\documentstyle[preprint,eqsecnum,aps]{revtex}

\begin{document}
\author{Jian-Qi Shen$,$ Hong-Yi Zhu$,$ and Pan Chen}
\address{State Key Laboratory of Modern Optical Instrumentation, Center for Optical\\
and Electromagnetic Research, College of Information Science and Engineering%
\\
Zhejiang Institute of Modern Physics and Department of Physics,\\
Zhejiang University, Hangzhou 310027, People$^{,}$s Republic of China}
\date{\today}
\title{Exact solutions of time-dependent three-generator systems}
\maketitle

\begin{abstract}
There exist a number of typical and interesting systems or models which
possess three-generator Lie-algebraic structure in atomic physics, quantum
optics, nuclear physics and laser physics. The well-known fact that all
simple 3-generator algebras are either isomorphic to the algebra $sl(2,C)$
or to one of its real forms enables us to treat these time-dependent quantum
systems in a unified way. By making use of the Lewis-Riesenfeld invariant
theory and the invariant-related unitary transformation formulation, the
present paper obtains exact solutions of the time-dependent Schr\"{o}dinger
equations governing various three-generator quantum systems. For some
quantum systems whose time-dependent Hamiltonians have no quasialgebraic
structures, we show that the exact solutions can also be obtained by working
in a sub-Hilbert-space corresponding to a particular eigenvalue of the
conserved generator (i.e., the time-independent invariant that commutes with
the time-dependent Hamiltonian). The topological property of geometric phase
factors in time-dependent systems is briefly discussed.
\end{abstract}

\section{Introduction}

Exact solutions and geometric phase factor\cite{Berry,Vinet} of
time-dependent spin model have been extensively investigated by many
authors. Bouchiat and Gibbons discussed the geometric phase for the spin-$1$
system\cite{Bouchiat}. Datta et al found the exact solution for the spin- $%
\frac{1}{2}$ system \cite{Datta} by means of the classical Lewis-Riesenfeld
theory, and Mizrahi calculated A-A phase for the spin- $\frac{1}{2}$ system 
\cite{Mizrahi} in a time-dependent magnetic field. The more systematic
approach to obtaining the formally exact solutions for the spin- $j$ system
was proposed by Gao et al\cite{Gao0} who made use of the Lewis-Riesenfeld
quantum theory\cite{Lewis}. In this spin- $j$ system, the three generators
of the Hamiltonian satisfy the commutation relations of $SU(2)$ Lie algebra.
In addition to the spin model, there exist many quantum systems whose
Hamiltonian is also constructed in terms of three generators of various Lie
algebras which we will illustrate in the following.

The invariant theory which can be better applied to solutions of the
time-dependent Schr\"{o}dinger equation was first proposed by Lewis and
Riesenfeld in 1969\cite{Lewis}. This theory is very appropriate for treating
the geometric phase factor. In 1991, Gao et al generalized this theory and
put forward the invariant-related unitary transformation method\cite{Gao1}.
Exact solutions for time-dependent systems obtained by using the generalized
invariant theory contain the geometric and dynamical phase\cite
{Gao2,Gao3,Gao4}. This formulation was developed from the Lewis-Riesenfeld$%
^{,}$s formal theory into a powerful tool for treating exact solutions of
the time-dependent Schr\"{o}dinger equation and geometric phase factor. In
the present paper, we obtain exact solutions of various time-dependent
three-generator systems based on these invariant theories.

This paper is organized as follows: In section 2, we set out several quantum
systems or models which have three-generator algebraic structure. In section
3, use is made of the invariant theories and exact solutions of various
time-dependent three-generator systems are obtained. In section 4, there are
some discussions concerning the closure property of the Lie algebraic
generators in the sub-Hilbert-space. In section 5, we conclude with some
remarks.

\section{The algebraic structures of various three-generator systems}

In our previous work\cite{Shen} we showed that the time-dependent
Schr\"{o}dinger equation is solvable if its Hamiltonian is constructed in
terms of the generators of Lie algebra. Then analyzing the algebraic
structures of Hamiltonians plays significant role in obtaining exact
solutions of the time -dependent systems. To our knowledge, a large number
of quantum systems which have three-generator Hamiltonians have been
investigated in the literature. In the following we set out these systems
and discuss the algebraic structures of their Hamiltonians.

(1) Spin model. The time evolution of the wavefunction of a spinning
particle in a magnetic field was studied by regarding it as a spin model\cite
{Mizrahi} whose Hamiltonian can be written

\begin{equation}
H(t)=c_{0}\{\frac{1}{2}\sin \theta \exp [-i\varphi ]J_{+}+\frac{1}{2}\sin
\theta \exp [i\varphi ]J_{-}+\cos \theta J_{3}\}  \label{eq1}
\end{equation}
with $J_{\pm }=J_{1}\pm iJ_{2}$ satisfying the commutation relations $%
[J_{3},J_{\pm }]=\pm J_{\pm },[J_{+},J_{-}]=2J_{3}.$ Analogous to this case,
in the gravitational theory of general relativity both spin-gravitomagnetic
interaction\cite{Kleinert} and spin-rotation coupling\cite
{Mashhoon1,Mashhoon2,Shen} are proved spin models. It can be verified that
the investigation of the propagation of a photon inside the noncoplanarly
curved optical fiber \cite{Chiao,Tomita,Kwiat} is also equivalent to that of
a spin model. The Hamiltonian of spin model is composed of three generators
which constitute $SU(2)$ algebra.

(2) Two-coupled harmonic oscillator. The Hamiltonian of the two-coupled
harmonic oscillator is of the form ( in the units $\hbar =1$)

\begin{equation}
H=\omega _{1}a_{1}^{\dagger }a_{1}+\omega _{2}a_{2}^{\dagger
}a_{2}+ga_{1}^{\dagger }a_{2}+g^{\ast }a_{2}^{\dagger }a_{1},  \label{eq2}
\end{equation}
where $a_{1}^{\dagger },a_{2}^{\dagger },a_{1},a_{2}$ are the creation and
annihilation operators for these two harmonic oscillators, respectively; $g$
and $g^{\ast }$ are the coupling coefficients. Set $J_{+}=a_{1}^{\dagger
}a_{2},J_{-}=a_{2}^{\dagger }a_{1},J_{3}=\frac{1}{2}(a_{1}^{\dagger
}a_{1}-a_{2}^{\dagger }a_{2}),N=\frac{1}{2}(a_{1}^{\dagger
}a_{1}+a_{2}^{\dagger }a_{2}),$ then we can show that the generators of this
Hamiltonian satisfy the commutation relations of $SU(2)$ algebra. Since $N$
commutes with $H$, i.e., $[N,H]=0,$we consequently say $N$ is an invariant
(namely, it is a conserved generator whose eigenvalue is time-independent).
In terms of $J_{\pm },J_{3}$ and $N,$ the Hamiltonian in the expression (\ref
{eq2}) can be rewritten as follows

\begin{equation}
H=\omega _{1}(N+J_{3})+\omega _{2}(N-J_{3})+gJ_{+}+g^{\ast }J_{-}.
\label{eq3}
\end{equation}
Another interesting Hamiltonian of the two-coupled harmonic oscillator is
written in the form

\begin{equation}
H=\omega _{1}a_{1}^{\dagger }a_{1}+\omega _{2}a_{2}^{\dagger
}a_{2}+ga_{1}a_{2}+g^{\ast }a_{1}^{\dagger }a_{2}^{\dagger }.  \label{eq4}
\end{equation}
If we take $K_{+}=a_{1}^{\dagger }a_{2}^{\dagger },K_{-}=a_{1}a_{2},K_{3}=%
\frac{1}{2}(a_{1}a_{1}^{\dagger }+a_{2}^{\dagger }a_{2}),N=\frac{1}{2}%
(a_{1}^{\dagger }a_{1}-a_{2}^{\dagger }a_{2}),$ the generators of the $%
SU(1,1)$ group are thus realized. The commutation relations are immediately
inferred as

\begin{equation}
\lbrack K_{3},K_{\pm }]=\pm K_{\pm },\quad [K_{+},K_{-}]=-2K_{3}.
\label{eq5}
\end{equation}

(3) $SU(1,1)\uplus _{s}h(4)$ Lie-algebra system. A good number of quantum
systems whose Hamiltonian is some combinations of the generators of a Lie
algebra, e.g., $SU(1,1)\uplus _{s}h(4)$ ($\uplus _{s}$denotes a semidirect
sum)\cite{Dattoli1,Dattoli2} which is used to discuss both the
non-Poissonian effects in a laser-plasma scattering and the pulse
propagation in a free-electron laser\cite{Dattoli2}. The $SU(1,1)\uplus
_{s}h(4)$ Hamiltonian is

\begin{equation}
H=AK_{3}+FK_{+}+F^{\ast }K_{-}+Ba^{\dagger }+B^{\ast }a+G,  \label{eq6}
\end{equation}
where $a$ and $a^{\dagger }$ are harmonic-oscillator annihilation and
creation operators, respectively.

(4) General harmonic oscillator. The Hamiltonian of the general harmonic
oscillator is given by\cite{Gao1}

\begin{equation}
H=\frac{1}{2}[Xq^{2}+Y(qp+pq)+Zp^{2}]+Fq,  \label{eq7}
\end{equation}
where the canonical coordinate $q$ and the canonical momentum $p$ satisfy
the commutation relation $[q,p]=i.$ The following three-generator Lie
algebra is easily derived

\begin{equation}
\lbrack q^{2},p^{2}]=2\{i(qp+pq)\},\quad [i(qp+pq),q^{2}]=4q^{2},\quad
[i(qp+pq),p^{2}]=-4p^{2}.  \label{eq8}
\end{equation}

(5) Charged particle moving in a magnetic field. The motion of a particle
with mass $\mu $ and charge $e$ in a homogeneous magnetic field $\vec{B}%
=(0,0,B)$ is described by the following Hamiltonian in the spherical
coordinates

\begin{equation}
H=-\frac{1}{2\mu }(\frac{\partial ^{2}}{\partial r^{2}}+\frac{2}{r}\frac{%
\partial }{\partial r}-\frac{1}{r^{2}}L^{2})+\frac{1}{8}\mu \omega ^{2}r^{2}-%
\frac{\omega }{2}L_{z}  \label{eq9}
\end{equation}
with $\omega =\frac{B}{\mu }.$ Since both $L_{z}$ and $L^{2}$ commute with $%
H $ and thus they are called invariants, only the operators associated with $%
r$ should be taken into consideration. We can show that if the following
operators are defined

\begin{equation}
K_{1}=\mu r^{2},\quad K_{2}=-\frac{1}{\mu }(\frac{\partial ^{2}}{\partial
r^{2}}+\frac{2}{r}\frac{\partial }{\partial r}-\frac{1}{r^{2}}L^{2}),\quad
K_{3}=-2i(\frac{3}{2}+r\frac{\partial }{\partial r}),  \label{eq10}
\end{equation}
$K_{1},K_{2}$ and $K_{3}$ form an algebra

\begin{equation}
\lbrack K_{1},K_{2}]=2iK_{3},\quad [K_{3},K_{2}]=4iK_{2},\quad
[K_{3},K_{1}]=-4iK_{1}.  \label{11}
\end{equation}
Apparently, $H$ can be rewritten in terms of the generators of this Lie
algebra.

(6) Two-level atomic coupling. The model under consideration is consisted of
two-level atom driven by the photons field\cite{Zhou}. The interaction part
of the Hamiltonian contains the transition operator $\left| 1\right\rangle
\left\langle 2\right| $ and $\left| 2\right\rangle \left\langle 1\right| $,
where $\left| 1\right\rangle $ and $\left| 2\right\rangle $ are the atomic
operators of the two-level atom. Simple calculation yields

\begin{eqnarray}
\lbrack \left| 1\right\rangle \left\langle 1\right| -\left| 2\right\rangle
\left\langle 2\right| ,\left| 1\right\rangle \left\langle 2\right| ]
&=&2\left| 1\right\rangle \left\langle 2\right| ,\quad [\left|
1\right\rangle \left\langle 1\right| -\left| 2\right\rangle \left\langle
2\right| ,\left| 2\right\rangle \left\langle 1\right| ]=-2\left|
2\right\rangle \left\langle 1\right| ,  \nonumber \\
\lbrack \left| 1\right\rangle \left\langle 2\right| ,\left| 2\right\rangle
\left\langle 1\right| ] &=&\left| 1\right\rangle \left\langle 1\right|
-\left| 2\right\rangle \left\langle 2\right| ,  \label{eq12}
\end{eqnarray}
which unfolds that the Hamiltonian contains a $SU(2)$ algebraic structure.

(7) Supersymmetric Jaynes-Cummings model. In addition to the ordinary
Jaynes-Cummings models\cite{Jaynes}, there exists a two-level multiphoton
Jaynes-Cummings model which possesses supersymmetric structure. In this
generalization of the Jaynes-Cummings model, the atomic transitions are
mediated by $k$ photons \cite{Sukumar,Sukumar2,Kien}. Singh has shown that
this model can be used to study multiple atom scattering of radiation and
multiphoton emission, absorption, and laser processes\cite{Singh}. The
Hamiltonian of this model under the rotating wave approximation is given by

\begin{equation}
H(t)=\omega (t)a^{\dagger }a+\frac{\omega _{0}(t)}{2}\sigma
_{z}+g(t)(a^{\dagger })^{k}\sigma _{-}+g^{\ast }(t)a^{k}\sigma _{+},
\label{eq13}
\end{equation}
where $a^{\dagger }$ and $a$ are the creation and annihilation operators for
the electromagnetic field, and obey the commutation relation $\left[
a,a^{\dagger }\right] =1$; $\sigma _{\pm }$ and $\sigma _{z}$ denote the
two-level atom operators which satisfy the commutation relation $\left[
\sigma _{z},\sigma _{\pm }\right] =\pm 2\sigma _{\pm }$ . We can verify that
this model is solvable and the complete set of exact solutions can be found
by working in a sub-Hilbert-space corresponding to a particular eigenvalue
of the supersymmetric generator $N^{^{\prime }}$%
\begin{equation}
N^{^{\prime }}=\left( 
\begin{array}{cc}
a^{k}(a^{\dagger })^{k} & 0 \\ 
0 & (a^{\dagger })^{k}a^{k}
\end{array}
\right) .  \label{eq14}
\end{equation}
It can be verified that $N^{^{\prime }}$ commutes with the Hamiltonian in (%
\ref{eq13}), $N^{^{\prime }}$ is therefore called the time-independent
invariant. Vogel and Welsch have studied the $k$-photon Jaynes-Cummings
model with coherent atomic preparation which is time-independent\cite{Vogel}%
. In the framework of the formulation presented in this paper, we can study
the totally time-dependent cases of work done by Vogel and Welsch.

(8) Two-level atom interacting with a generalized cavity. Consider the
following Hamiltonian\cite{Yu}

\begin{equation}
H=r(A_{0})+s(A_{0})\sigma _{z}+gA_{-}\sigma _{+}+g^{\ast }A_{+}\sigma _{-}
\label{eq15}
\end{equation}
where $r(A_{0})$ and $s(A_{0})$ are well-defined real functions of $A_{0},$
and $A_{0},A_{\pm }$ satisfy the commutation relations $[A_{0},A_{\pm }]=\pm
mA_{\pm }\cite{Bonatsos}$. One can show that this Hamiltonian possesses a
three-generator algebraic structure.

(9) The interaction between a hydrogenlike atom and an external magnetic
field. This model is described by

\begin{eqnarray}
H &=&\alpha \vec{L}\cdot \vec{S}+\beta (L_{z}+2S_{z})  \nonumber \\
&=&\beta L_{z}+(\frac{1}{2}\alpha L_{z}+\beta )\sigma _{z}+\frac{1}{2}\alpha
(L_{-}\sigma _{+}+L_{+}\sigma _{-})  \label{eq16}
\end{eqnarray}
with $L_{\pm }$ $=L_{x}\pm iL_{y}.$ It is evidently seen that this form of
Hamiltonian is analogous to that in (\ref{eq15}).

(10) Coupled two-photon lasers. The Hamiltonian of this model is in fact the
combination of that of the two-coupled harmonic oscillator and general
harmonic oscillator. One can show that there exists a $SU(2)$ algebraic
structure in this model\cite{Gomes}.

From what has been discussed above we can draw a conclusion that a number of
typical and useful systems or models in laser physics, atomic physics and
quantum optics can be attributed to various three-generator type. The
analysis of these algebraic structures shows the solvability of these
systems. It should be noted that, in the literature, most of above systems
or models are only investigated in the stationary cases where the
coefficients of the Hamiltonians are totally time-independent. Some systems
are investigated for the Hamiltonian with partly time-dependent
coefficients. In the present paper, we will give exact solutions of the
time-dependent Schr\"{o}dinger equation of all these systems or models where
the coefficients of the Hamiltonians are totally time-dependent.

\section{ Exact solutions of time-dependent Schr\"{o}dinger equation}

Time evolution of most above systems or models is governed by the
Schr\"{o}dinger equation

\begin{equation}
i\frac{\partial \left| \Psi (t)\right\rangle _{s}}{\partial t}=H(t)\left|
\Psi (t)\right\rangle _{s},  \label{eq17}
\end{equation}
where the Hamiltonian is constructed by three generators $A,B$ and $C$ and
is often given as follows

\begin{equation}
H(t)=\omega (t)\{\frac{1}{2}\sin \theta (t)\exp [-i\phi (t)]A+\frac{1}{2}%
\sin \theta (t)\exp [i\phi (t)]B+\cos \theta (t)C\}  \label{eq18}
\end{equation}
with $A,B$ and $C$ satisfying the general commutation relations of a Lie
algebra

\begin{equation}
\lbrack A,B]=nC,\quad \lbrack C,A]=mA,\quad \lbrack C,B]=-mB,  \label{eq19}
\end{equation}
where $m$ and $n$ are the structure constants of this Lie algebra. Since all
simple $3$-generator algebras are either isomorphic to the algebra $sl(2,C)$
or to one of its real forms, we treat these time-dependent quantum systems
in a unified way. According to the Lewis-Riesenfeld invariant theory, an
operator $I(t)$ that agrees with the following invariant equation\cite{Lewis}

\begin{equation}
\frac{\partial I(t)}{\partial t}+\frac{1}{i}[I(t),H(t)]=0  \label{eq20}
\end{equation}
is called an invariant whose eigenvalue is time-independent, i.e.,

\begin{equation}
I(t)\left| \lambda ,t\right\rangle _{I}=\lambda \left| \lambda
,t\right\rangle _{I},\quad \frac{\partial \lambda }{\partial t}=0.
\label{eq21}
\end{equation}
It is seen from Eq. (\ref{eq20}) that $I(t)$ is the linear combination of $%
A,B$ and $C$\bigskip\ and may be generally written\qquad

\begin{equation}
I(t)=y\{\frac{1}{2}\sin a(t)\exp [-ib(t)]A+\frac{1}{2}\sin a(t)\exp
[ib(t)]B\}+\cos a(t)C,  \label{eq22}
\end{equation}
where the constant $y$ will be determined below. It should be pointed out
that it is not the only way to construct the invariants. Since the product
of two invariants also satisfies Eq. (\ref{eq20})\cite{Gao1}, there are
infinite invariants of a time-dependent quantum system. But the form in Eq. (%
\ref{eq22}) is the most convenient and useful one. Substitution of (\ref
{eq22}) into Eq.(\ref{eq20}) yields

\begin{eqnarray}
y\exp (-ib)(\dot{a}\cos a-i\dot{b}\sin a)-im\omega \lbrack \exp (-i\phi
)\cos a\sin \theta -y\exp (-ib)\sin a\cos \theta ] &=&0,  \nonumber \\
\dot{a}+\frac{ny}{2}\omega \sin \theta \sin (b-\phi ) &=&0.  \label{eq23}
\end{eqnarray}
where dot denotes the time derivative. The time-dependent parameters $a$ and 
$b$ are determined by these two auxiliary equations.

It is easy to verify that the particular solution $\left| \Psi
(t)\right\rangle _{s}$ of the Schr\"{o}dinger equation can be expressed in
terms of the eigenstate $\left| \lambda ,t\right\rangle _{I}$ of the
invariant $I(t),$ namely,

\begin{equation}
\left| \Psi (t)\right\rangle _{s}=\exp [\frac{1}{i}\varphi (t)]\left|
\lambda ,t\right\rangle _{I}  \label{eq24}
\end{equation}
with 
\begin{equation}
\varphi (t)=\int_{0I}^{t}\left\langle \lambda ,t^{^{\prime }}\right|
[H(t^{^{\prime }})-i\frac{\partial }{\partial t^{^{\prime }}}]\left| \lambda
,t^{^{\prime }}\right\rangle _{I}dt^{^{\prime }}.
\end{equation}
The physical meanings of $\int_{0I}^{t}\left\langle \lambda ,t^{^{\prime
}}\right| H(t^{^{\prime }})\left| \lambda ,t^{^{\prime }}\right\rangle
_{I}dt^{^{\prime }}$ and $\int_{0I}^{t}\left\langle \lambda ,t^{^{\prime
}}\right| -i\frac{\partial }{\partial t^{^{\prime }}}\left| \lambda
,t^{^{\prime }}\right\rangle _{I}dt^{^{\prime }}$ are dynamical and
geometric phase, respectively.

Since the expression (\ref{eq24}) is merely a formal solution of the
Schr\"{o}dinger equation, in order to get the explicit solutions we make use
of the invariant-related unitary transformation formulation\cite{Gao1} which
enables one to obtain the complete set of exact solutions of the
time-dependent Schr\"{o}dinger equation (\ref{eq17}). In accordance with the
invariant-related unitary transformation method, the time-dependent unitary
transformation operator is often of the form

\begin{equation}
V(t)=\exp [\beta (t)A-\beta ^{\ast }(t)B]  \label{eq25}
\end{equation}
with $\beta (t)=-\frac{a(t)}{2}x\exp [-ib(t)],\quad \beta ^{\ast }(t)=-\frac{%
a(t)}{2}x\exp [ib(t)].$ By making use of the Glauber formulae, lengthy
calculation yields

\begin{eqnarray}
I_{V} &=&V^{\dagger }(t)I(t)V(t)=\{\frac{y}{2}\exp (-ib)\sin a\cos [(\frac{mn%
}{2})^{\frac{1}{2}}ax]-\frac{(\frac{mn}{2})^{\frac{1}{2}}}{n}\exp (-ib)\cos
a\sin [(\frac{mn}{2})^{\frac{1}{2}}ax]\}A  \nonumber \\
&&+\{\frac{y}{2}\exp (ib)\sin a\cos [(\frac{mn}{2})^{\frac{1}{2}}ax]-\frac{(%
\frac{mn}{2})^{\frac{1}{2}}}{n}\exp (ib)\cos a\sin [(\frac{mn}{2})^{\frac{1}{%
2}}ax]\}B  \nonumber \\
&&+\{\cos a\cos [(\frac{mn}{2})^{\frac{1}{2}}ax]+\frac{(\frac{mn}{2})^{\frac{%
1}{2}}}{m}y\sin a\sin [(\frac{mn}{2})^{\frac{1}{2}}ax]\}C.
\end{eqnarray}
It can be easily seen that when 
\begin{equation}
y=\frac{m}{(\frac{mn}{2})^{\frac{1}{2}}},\quad x=\frac{1}{(\frac{mn}{2})^{%
\frac{1}{2}}},
\end{equation}
one may derive that $I_{V}=C$ which is time-independent. Thus the
eigen-equation of the time-independent invariant $I_{V}$ may be written in
the form

\begin{equation}
I_{V}\left| \lambda \right\rangle =\lambda \left| \lambda \right\rangle
,\quad \left| \lambda \right\rangle =V^{\dagger }(t)\left| \lambda
,t\right\rangle _{I}.  \label{eq27}
\end{equation}
Under the transformation $V(t),$ the Hamiltonian $H(t)$ can be changed into

\begin{eqnarray}
H_{V}(t) &=&V^{\dagger }(t)H(t)V(t)-V^{\dagger }(t)i\frac{\partial V(t)}{%
\partial t}  \nonumber \\
&=&\{\omega \lbrack \cos a\cos \theta +\frac{(\frac{mn}{2})^{\frac{1}{2}}}{m}%
\sin a\sin \theta \cos (b-\phi )]+\frac{\dot{b}}{m}(1-\cos a)\}C
\label{eq28}
\end{eqnarray}
by the aid of Baker-Campbell-Hausdorff formula\cite{Wei}

\begin{equation}
V^{\dagger }(t)\frac{\partial }{\partial t}V(t)=\frac{\partial }{\partial t}%
L+\frac{1}{2!}[\frac{\partial }{\partial t}L,L]+\frac{1}{3!}[[\frac{\partial 
}{\partial t}L,L],L]+\frac{1}{4!}[[[\frac{\partial }{\partial t}%
L,L],L],L]+\cdots
\end{equation}
with $V(t)=\exp [L(t)].$ Hence, with the help of Eq.(\ref{eq24}) and Eq.(\ref
{eq27}), the particular solution of the Schr\"{o}dinger equation is obtained

\begin{equation}
\left| \Psi (t)\right\rangle _{s}=\exp [\frac{1}{i}\varphi (t)]V(t)\left|
\lambda \right\rangle  \label{eq29}
\end{equation}
with the phase

\begin{eqnarray}
\varphi (t) &=&\int_{0}^{t}\left\langle \lambda \right| [V^{\dagger
}(t^{^{\prime }})H(t^{^{\prime }})V(t^{^{\prime }})-V^{\dagger }(t^{^{\prime
}})i\frac{\partial }{\partial t^{^{\prime }}}V(t^{^{\prime }})]\left|
\lambda \right\rangle dt^{^{\prime }}=\varphi _{d}(t)+\varphi _{g}(t) 
\nonumber \\
&=&\lambda \int_{0}^{t}\{\omega \lbrack \cos a\cos \theta +\frac{(\frac{mn}{2%
})^{\frac{1}{2}}}{m}\sin a\sin \theta \cos (b-\phi )]+\frac{\dot{b}}{m}%
(1-\cos a)\}dt^{^{\prime }},  \label{eq30}
\end{eqnarray}
where the dynamical phase is $\varphi _{d}(t)=\lambda \int_{0}^{t}\omega
\lbrack \cos a\cos \theta +\frac{(\frac{mn}{2})^{\frac{1}{2}}}{m}\sin a\sin
\theta \cos (b-\phi )]dt^{^{\prime }}$ and the geometric phase is $\varphi
_{g}(t)=\lambda \int_{0}^{t}\frac{\dot{b}}{m}(1-\cos a)dt^{^{\prime }}.$ It
is seen that the former phase is in connection with the coefficients of the
Hamiltonian such as $\omega ,\cos \theta ,\sin \theta ,$ etc., whereas the
latter is not immediately related to these coefficients. If the parameter $a$
is taken to be time-independent, $\varphi _{g}(T)=\lambda \int_{0}^{T}\frac{%
\dot{b}}{m}(1-\cos a)dt^{^{\prime }}=\frac{\lambda }{m}[2\pi (1-\cos a)]$
where $2\pi (1-\cos a)$ is an expression for the solid angle over the
parameter space of the invariant. It is of interest that $\frac{\lambda }{m}%
[2\pi (1-\cos a)]$ is equal to the magnetic flux produced by a monopole of
strength $\frac{\lambda }{4\pi m}$ existing at the origin of the parameter
space. This, therefore, implies that geometric phase differs from dynamical
phase and it involves the global and topological properties of the time
evolution of a quantum system. This fact indicates the geometric or
topological meaning of $\varphi _{g}(t).$

The expression (\ref{eq29}) is a particular exact solution corresponding to $%
\lambda $ and the general solutions of the time-dependent Schr\"{o}dinger
equation are easily obtained by using the linear combinations of all these
particular solutions. Generally speaking, in Quantum Mechanics, solution
with chronological-product operator (time-order operator) $P$ is often
called the formal solution. In the present paper, however, the solution of
the Schr\"{o}dinger equation governing a time-dependent system is sometimes
termed the explicit solution, for reasons of the fact that it does not
involve time-order operator. But, on the other hand, by using
Lewis-Riesenfeld invariant theory, there always exist time-dependent
parameters, for instance, $a(t)$ and $b(t)$ in this paper which are
determined by the auxiliary equations (\ref{eq23}). In the traditional
practice, when employed in experimental analysis and compared with
experimental results, these nonlinear auxiliary equations should be solved
often by means of numerical calculation. From above viewpoints, the concept
of explicit solution is understood in a relative sense, namely, it can be
considered explicit solution when compared with the time-evolution operator $%
U(t)=P\exp [\frac{1}{i}\int_{0}^{t}H(t^{^{\prime }})dt^{^{\prime }}]$
involving time-order operator, $P$; whereas, it cannot be considered
completely explicit solution for it is expressed in terms of some
time-dependent parameters which should be obtained via the auxiliary
equations. Hence, conservatively speaking, we regard the solution of the
time-dependent system presented in the paper as exact solution rather than
explicit solution.

\section{Discussions\qquad}

In the previous section, we obtain exact solutions of some time-dependent
three-generator systems or models by using these invariant theories. In what
follows there is the closure property of the Lie algebraic generators in the
sub-Hilbert-space that should be discussed.

The generalized invariant theory can only be applied to the study of the
system for which there exists the quasialgebra defined in Ref.\cite{Mizrahi}%
. It is easily seen from (\ref{eq15}) that there is no such quasialgebra for
the Hamiltonian (\ref{eq15}) of example (7)(supersymmetric Jaynes-Cummings
model) and example (8)(two-level atom interacting with a generalized
cavity). We generalize the method which has been used for finding the
dynamical algebra $O(4)$ of the hydrogen atom to treat this type of
time-dependent models. In the case of hydrogen, the dynamical algebra $O(4)$
was found by working in the sub-Hilbert-space corresponding to a particular
eigenvalue of the Hamiltonian \cite{Schiff}. In this paper, we will show
that a generalized quasialgebra can also be found by working in a
sub-Hilbert-space corresponding to a particular eigenvalue of $\Delta
=A_{0}+m\frac{1+\sigma _{z}}{2}$ in the time-dependent model of two-level
atom interacting with a generalized cavity. This generalized quasialgebra
enables one to obtain the complete set of exact solutions for the
Schr\"{o}dinger equation. It is easily verified that $\Delta $ commutes with 
$H(t)$ and is a time-independent invariant according to Eq. (\ref{eq20}).
Then in order to unfold the algebraic structure of the Hamiltonian (\ref
{eq15}), the following three operators are defined\cite{Yu}

\begin{equation}
\Sigma _{1}=\frac{1}{2[\chi (\Delta )]^{\frac{1}{2}}}(A_{-}\sigma
_{+}+A_{+}\sigma _{-}),\quad \Sigma _{2}=\frac{i}{2[\chi (\Delta )]^{\frac{1%
}{2}}}(A_{+}\sigma _{-}-A_{-}\sigma _{+}),\quad \Sigma _{3}=\frac{1}{2}%
\sigma _{z},  \label{eq34}
\end{equation}
where $\chi =\left\langle n\right| A_{+}A_{-}\left| n\right\rangle ,\left|
n\right\rangle $ denotes the eigenstates of $A_{0}.$ It is easy to see all
these operators commute with $\Delta $ and the quasialgebra $\{H,\Sigma
_{1},\Sigma _{2},\Sigma _{3}\}$ is thus found. Therefore, this type of
time-dependent models is proved solvable by working in a sub-Hilbert-space
corresponding to the eigenstates of the time-independent invariant. \qquad

For the supersymmetric Jaynes-Cummings model, the commutation relations of
its supersymmetric Lie- algebraic structure are

\begin{eqnarray}
\left[ Q^{\dagger },Q\right] &=&\lambda \sigma _{z},\quad \left[ N,Q\right]
=Q,\quad \left[ N,Q^{\dagger }\right] =-Q^{\dagger },  \nonumber \\
\left[ Q,\sigma _{z}\right] &=&2Q,\quad \left[ Q^{\dagger },\sigma _{z}%
\right] =-2Q^{\dagger },  \label{eq35}
\end{eqnarray}
where

\begin{equation}
N=a^{\dagger }a+\frac{k-1}{2}\sigma _{z}+\frac{1}{2},\quad Q=(a^{\dagger
})^{k}\sigma _{-},\quad Q^{\dagger }=a^{k}\sigma _{+},  \label{eq36}
\end{equation}
and $\lambda $ denotes the eigenvalue of the time-independent invariant $%
N^{^{\prime }}.$ By the aid of (\ref{eq35}) and (\ref{eq36}), the
Hamiltonian (\ref{eq13}) of this supersymmetric Jaynes-Cummings model can be
rewritten as

\begin{equation}
H(t)=\omega (t)N+\frac{\omega (t)-\delta (t)}{2}\sigma _{z}+g(t)Q+g^{\ast
}(t)Q^{\dagger }-\frac{\omega (t)}{2}.  \label{eq37}
\end{equation}
Then this time-dependent model can be exactly solved by using the
invariant-related unitary transformation formulation where the unitary
transformation operator is of the form

\begin{equation}
V(t)=\exp [\beta (t)Q-\beta ^{\ast }(t)Q^{\dagger }].  \label{eq38}
\end{equation}

It should be noted that the above approach to the time-dependent
Jaynes-Cummings model is also appropriate for treating the periodic decay
and revival of some multiphoton-transitions models which has been
investigated by Sukumar and Buck\cite{Sukumar2}.

It is readily verified that in the sub-Hilbert-space corresponding to a
particular eigenvalue of the conserved generator (the time-independent
invariant), the Hamiltonian of original two-level Jaynes-Cummings model
(mono-photon case)\cite{Jaynes} possesses the $SU(2)$ Lie-algebraic
structure, and three-level two-mode mono-photon model possesses the $SU(3)$
structure. The solution of the time-dependent case of $SU(2)$
Jaynes-Cummings model is easily obtained by taking the number of photons
mediating in the process of atomic transitions $k=1.$ Since Shumovsky et al
have considered the three-level two-mode multiphoton Jaynes-Cummings model 
\cite{Shumovsky} whose Hamiltonian is time-independent, it is also of
interest to exactly solve the time-dependent supersymmetric three-level
two-mode multiphoton Jaynes-Cummings model by means of invariant theories.

\section{Summary}

On the basis of the fact that all simple 3-generator algebras are either
isomorphic to the algebra $sl(2,C)$ or to one of its real forms, exact
solutions of the time-dependent Schr\"{o}dinger equation of all
three-generator systems or models in quantum optics, nuclear physics, solid
state physics, molecular and atomic physics and laser physics are offered by
making use of the Lewis-Riesenfeld invariant theory and the
invariant-related unitary transformation formulation in the present paper.
Since it appears only in systems with time-dependent Hamiltonian, the
geometric phase factor would be easily investigated if the exact solutions
of time-dependent systems had been obtained. In view of above discussions,
the invariant-related unitary transformation formulation is a useful tool
for treating the geometric phase factor and the time-dependent
Schr\"{o}dinger equation. This formulation replaces the eigenstates of the
time-dependent invariants with those of the time-independent invariants
through the unitary transformation. Apparently, it is also applicable to the
algebraic structure whose number of generators is more than three.
Additionally, it should be pointed out that the time-dependent
Schr\"{o}dinger equation is often investigated in the literature, whereas
less attention is paid to the time-dependent Klein-Gordon equation. Work in
this direction is under consideration and will be published elsewhere.

Acknowledgments This project is supported by the National Natural Science
Foundation of China under the project No.$19775040$ and $30000034$. The
authors thank Prof. Gao Xiao-Chun for offering the knowledge concerning the
closure property of the Lie algebraic generators in sub-Hilbert-space.

\end{document}